\documentclass[a4paper,twocolumn,11pt]{quantumarticle}
\pdfoutput=1
\usepackage[utf8]{inputenc}
\usepackage[english]{babel}
\usepackage[T1]{fontenc}
\usepackage{amsmath}
\usepackage{hyperref}
\usepackage{graphicx,bm}
\usepackage{xcolor}

\usepackage{tikz}
\usepackage[autostyle]{csquotes}

\begin{document}

\title{Theory of temporal three-photon interference}

\author{Nilakshi Senapati}
\affiliation{Department of Physics, Indian Institute of Technology Kanpur, Kanpur - 208016, UP, India}
\orcid{0009-0008-8407-2883}
\author{Girish Kulkarni}
\affiliation{Department of Physics, Indian Institute of Technology Ropar, Rupnagar - 140001, Punjab, India}
\orcid{0000-0001-5847-1612}
\author{Anand K. Jha}
\affiliation{Department of Physics, Indian Institute of Technology Kanpur, Kanpur - 208016, UP, India}
\email{akjha@iitk.ac.in}
\orcid{0000-0002-5186-6264}
\maketitle

\begin{abstract}
One-photon interference is best described by Dirac's famous dictum that ``Each photon interferes only with itself.'' A generic two-alternative one-photon interference is characterized in terms of a single parameter, namely, the path-length difference between the two alternatives. Similarly, in the context of entangled two-photon fields produced by parametric down-conversion (PDC), two-photon interference can be understood as ``each two-photon interfering only with itself.'' A generic two-alternative two-photon interference is characterized in terms of two independent length parameters: one is called the two-photon path-length difference, which plays a role analogous to the path-length difference in one-photon interference, and the other one is called the two-photon path-asymmetry-length difference. The second parameter, which has no counterpart in one-photon interference, governs manifestly nonclassical effects such as the Hong-Ou-Mandel (HOM) effect. The recent demonstrations of cascaded PDC (CPDC) and the hopeful prospects of realizing third-order PDC (TOPDC) for the generation of three-photon entangled states are paving the way for experimental studies on genuine three-photon interference. In this article, we formulate three-photon interference in terms of ``each three-photon interfering only with itself.'' We show that although a generalized two-alternative three-photon interference setup based on CPDC or TOPDC involves eight different length parameters, the interference can be fully characterized in terms of only three independent parameters. The first parameter is the three-photon path-length difference, which has a direct analog in the one-photon and two-photon cases, and the other two parameters quantify the path-asymmetry length. Unlike two-photon interference, which requires only one parameter to quantify path-asymmetry, \textit{two} independent parameters are needed in three-photon interference. This results in a broader class of nonclassical three-photon effects, including three-photon HOM-type effects. Our work provides the theoretical basis for existing and future three-photon interference experiments exploring the rich and complex quantum correlations associated with three-particle entanglement and potentially enabling the development of novel protocols for harnessing those correlations.
\end{abstract}

\section{Introduction}

Feynman had famously remarked that interference contains the only mystery of quantum mechanics \cite{feynman2011book}. The study of single-photon interference dates back to Thomas Young’s classic double-slit experiment  \cite{born&wolf1999, mandel&wolf1995}, while the study of multi-photon interference effects commences with the experiments of Hanbury-Brown and Twiss \cite{hanburyBrown1956, hanburyBrown1956a}. In a single-photon interference experiment, a single detector is used for measuring the probability of detecting a photon as a function of time or space, while in a multi-photon interference experiment, multiple detectors are used for measuring the joint probability of detection \cite{glauber1963pr, glauber1963pr2}. Interference effects with single-photon states are best described in terms of Dirac's famous dictum \cite{dirac1947} that ``Each photon interferes only with itself. Interference between two different photons never occurs.'' In his Nobel Lecture \cite{glauber2006rmp}, Glauber praised Dirac's dictum as being ``ringingly clear" and went on to describe its full import by noting that ``it is not the photons that interfere physically, it is their probability amplitudes that interfere—and probability amplitudes can be defined equally well for arbitrary numbers of photons." Although Dirac's description for one-photon interference has been extended to interference effects involving two-photon fields \cite{hong1985pra, franson1989prl, ou1989pra, grice1997pra,kim2000pra,jha2008pra,kulkarni2017josab, jha2010pra}, a comprehensive description is still pending for fields with three or more photons.

Quantum entanglement, which refers to the inseparability of the global quantum state of a composite multiparticle system into local states of the individual constituent particles, is a key feature of quantum theory that accounts for curious fundamental phenomena such as nonlocality \cite{einstein1935pr,bell1964physics} and useful practical applications such as teleportation \cite{bennett1993prl, bouwmeester1999prl, weihs1998prl, shalm2015prl}. At present, one of the most widely used experimental platform for generating entangled two-photon quantum states is parametric down-conversion (PDC) -- a second-order $\chi^{(2)}$ nonlinear optical process in which a single photon from an incident field known as the pump gets annihilated to produce two entangled photons termed as signal and idler \cite{boyd2020nlo}. The strong nonclassical correlations between the signal and idler photons manifest themselves quite strikingly through two-photon interference effects, wherein high-visibility interference fringes are observed in the two-photon coincidence rate, even in the absence of any modulation in the individual one-photon intensities \cite{mandel1999rmp}. For instance, the nonclassical temporal correlations of the signal and idler photons manifest themselves through temporal two-photon interference effects such as the Hong-Ou-Mandel (HOM) effect \cite{hong1987prl}, Franson effect \cite{franson1989prl}, induced coherence without induced emission \cite{zou1991prl, wang1991pra}, and frustrated two-photon creation \cite{herzog1994prl}. Such two-photon interference experiments have historically played a crucial role in advancing our fundamental understanding of two-particle entanglement and how it can be harnessed for quantum applications.

The theoretical description of temporal two-photon interference effects has been gradually developed over many years \cite{hong1985pra, franson1989prl, ou1989pra, grice1997pra,kim2000pra,jha2008pra,kulkarni2017josab}. It is now known that the temporal two-photon interference effects can be quantitatively explained within the ambit of a single theoretical formalism based on describing them in terms of \enquote{each two-photon interfering only with itself} \cite{jha2008pra,kulkarni2017josab}. Within this formalism, it has been shown that although a generic two-alternative two-photon interference setup has six different tunable length parameters, the temporal interference effects can be completely characterized in terms of only two independent length parameters. In other words, the two-photon temporal coherence function factorizes into two parts corresponding to each parameter. One of the length parameters is called the two-photon path-length difference, which plays a role analogous to the path-length difference parameter in one-photon interference. Thus, as a function of this parameter, one observes two-photon interference fringes with periodicity fixed by the central wavelength of the pump field, where the corresponding temporal coherence function is determined by the frequency correlations of the pump field. In essence, the theory shows that the pump's temporal coherence gets transferred to the generated two-photon field \cite{kulkarni2017josab}. The second parameter is called the two-photon path-asymmetry-length difference, which quantifies the path asymmetry between the two photons. The coherence function corresponding to this parameter is determined by the PDC phase-matching and depends on the frequency correlations of the down-converted photons. This parameter has no analog in one-photon interference and is responsible for exclusive two-photon effects. The Hong-Ou-Mandel (HOM) effect \cite{hong1987prl}, which is one of the most famous two-photon coherence effects, can be understood in terms of how two-photon coherence varies as a function of two-photon path-asymmetry-length difference in a setup that involves mixing of two down-converted photons at a beam splitter \cite{jha2008pra}. The above formalism has also been used to show that the pump's degree of temporal coherence sets an upper bound on the concurrence of the generated two-qubit energy-time entangled two-photon field \cite{kulkarni2017josab}. Thus, the formalism not only provides a unified description of all the temporal two-photon interference experiments but also reveals the fundamental origin of two-photon energy-time entanglement and shows how that entanglement can be controlled by tuning the pump's temporal coherence in various quantum applications \cite{jha2008prl, brendel1999prl,tittel2000prl,marcikic2004prl,pan2012rmp}.

\begin{figure}[t!]
	\centering
	\includegraphics[scale=0.45]{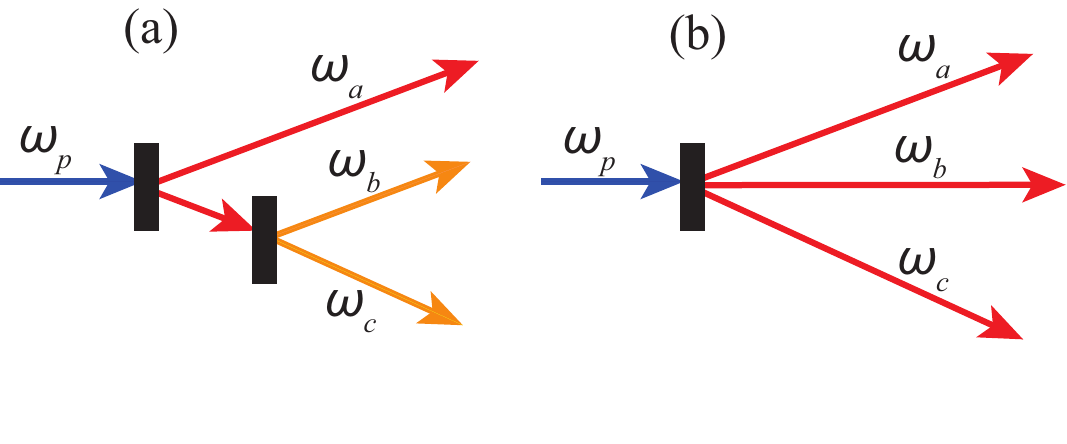}
	\caption{Schematic depictions of entangled three-photon state generation based on: (a) Cascaded parametric down-conversion (CPDC), (b) Third-order parametric down-conversion (TOPDC).}
	\label{fig:3photon}
\end{figure}

In recent years, researchers have been investigating entangled states involving three or more particles with the aim of exploring fundamental phenomena and technological applications beyond those associated with two-particle entangled states. For instance, theoretical studies have shown that three-particle entangled states exhibit peculiar phenomena such as entanglement monogamy \cite{coffman2000pra}, two inequivalent kinds of entanglement \cite{dur2000pra}, and incompatibility with local realism even at the level of single measurements \cite{greenberger1990ajp}, which have no counterparts in two-particle entangled states. Some theoretical studies have also investigated the continuous-variable three-photon quantum entanglement and its non-Gaussian character \cite{gonzalez2018prl, agusti2020prl, zhang2021pra}. On the experimental front, some early studies first demonstrated measurement of three-photon entangled states by combining photon pairs from different PDC processes using beam-splitters and then postselecting a part of the full state \cite{bouwmeester1999prl,eibl2004prl}. However,  such states cannot be used for tasks such as Bell state preparation \cite{zukowski1993prl}, which is known to be useful in quantum repeaters \cite{briegel1998prl}, loophole-free Bell tests \cite{cabello2012prx}, and optical quantum computing \cite{pittman2001pra,browne2005prl}. Furthermore, postselection also prevents the identification of conservation principles that govern the transfer of correlations from the pump to the generated entangled states \cite{kulkarni2016pra}. Consequently, researchers are turning towards non-postselective methods such as cascaded PDC (CPDC) and third-order PDC (TOPDC) for generating three-photon entangled states.

\begin{figure*}[t!]
	\centering
	\includegraphics[scale=0.7]{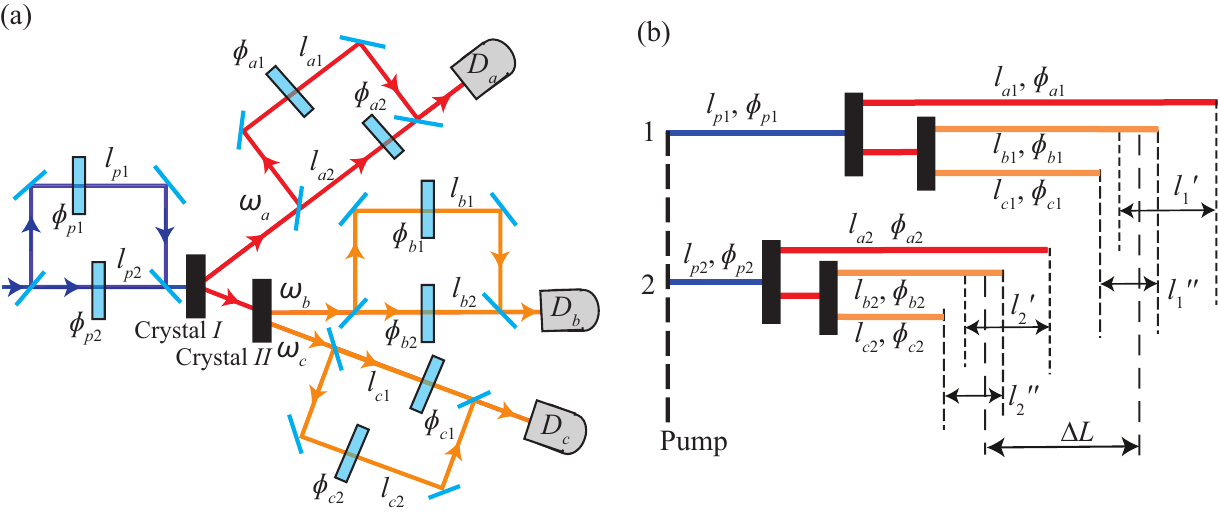}
	\caption{(a) Schematic setup of a temporal three-photon interference experiment with CPDC source, (b) Path diagram depicting the various photon paths in the experiment for CPDC case.}
	\label{fig2}
\end{figure*}

In CPDC, one daughter photon from the first PDC process pumps the second PDC process to effectively produce three entangled photons as shown in Fig. 1(a), whereas in TOPDC, one photon from an incident pump field is annihilated in a third-order ($\chi^{(3)}$) nonlinear interaction to produce three entangled daughter photons as shown in Fig. 1(b). TOPDC has been the subject of a number of theoretical studies over the last two decades \cite{felbinger1998prl,chekhova2005pra,corona2011pra,borschevskaya2015lpl,okoth2019pra,dominguez2020pra, cavanna2020pra, banic2022pra}, but an experimental generation of three-photon states has been realized only in the microwave domain using superconducting circuits \cite{chang2020prx}. At present, efforts are underway to experimentally realize TOPDC in the optical domain using bulk crystals and optical fibers with high third-order nonlinearities. A recent work \cite{Bacaoco25arxiv} has come up with a novel method for modelling TOPDC in dielectric nonlinear resonant metasurfaces which can potentially be used as free-space three-photon source. In contrast, three-photon state generation from CPDC has been realized using periodically-poled crystals and waveguides for over a decade now \cite{hubel2010nature, shalm2013natphys, hamel2014natphot, agne2017prl}. Although the generation efficiencies are admittedly extremely low with Ref. \cite{agne2017prl} reporting a triplet count rate of less than 200 per hour, the three-photon states generated using CPDC have been successfully employed for studying energy-time correlations and three-qubit polarization entanglement \cite{hubel2010nature, shalm2013natphys, hamel2014natphot}. More recently, CPDC has also been used to demonstrate genuine three-photon temporal interference \cite{agne2017prl} and OAM conservation at single-photon levels \cite{kopf2025prl}. Concurrently, another experimental study used three-photon entangled states generated by interfering heralded single photons on a fiber tritter and investigated three-photon interference as a function of the collective triad phase \cite{menssen2017prl}. As more such studies continue to become experimentally feasible, there is a need to formulate a comprehensive theoretical description of three-photon interference effects.

In this article, we formulate temporal three-photon interference in terms of a generalized version of Dirac's dictum, i.e, ``a three-photon interfering with itself'', for prototypical setups involving CPDC- and TOPDC-based sources. The article is organized as follows. In Sec.~ \ref{sec:sec2}, we present three-photon path diagrams of the interfering alternatives and define the three length parameters. In sections ~\ref{sec:Theory} and ~\ref{sec:effects}, we calculate the three-photon detection probabilities and describe the different categories of three-photon effects in terms of three independent parameters, respectively. In Sec.~\ref{conclusion}, we conclude with a summary and outlook of our study.

\section{Three-photon path diagram and the three length parameters}\label{sec:sec2}

\begin{figure*}[t!]
	\centering
	\includegraphics[scale=0.7]{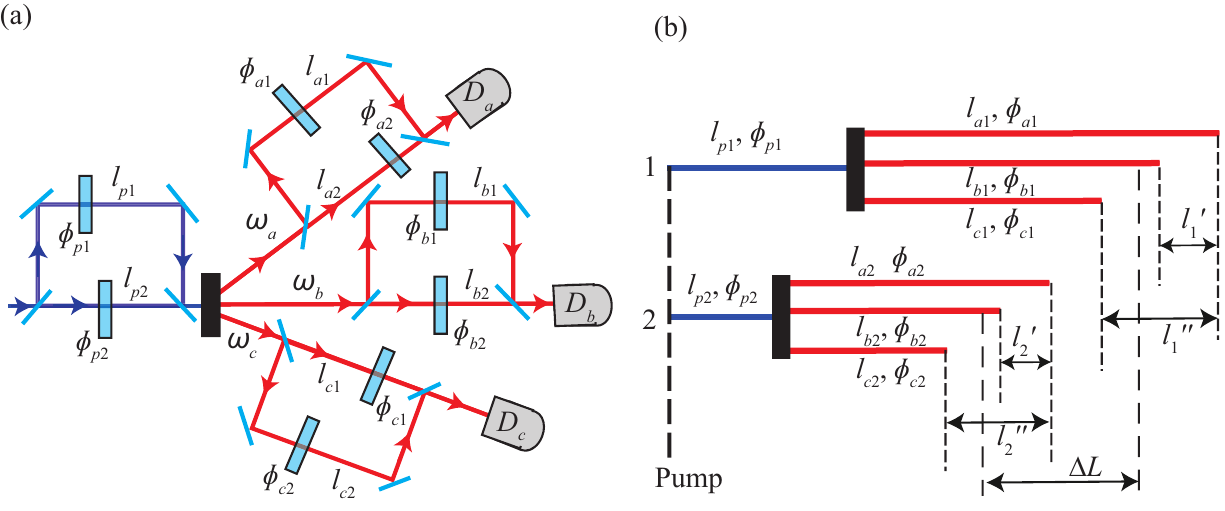}
	\caption{(a) Schematic setup of a temporal three-photon interference experiment with TOPDC source, (b) Path diagram depicting the various photon paths in the experiment for TOPDC case}
	\label{fig3}
\end{figure*}

Figures~\ref{fig2}(a) and \ref{fig3}(a) depict prototypical three-photon interference setups involving three-photon states generated from CPDC and TOPDC, respectively. The corresponding three-photon path diagrams have been depicted in Figs.~\ref{fig2}(b) and \ref{fig3}(b). In each of these setups, there are two different alternative pathways by which the three-photon state is generated and subsequently detected at the single-photon detectors $D_a$, $D_b$ and $D_c$, where the subscripts $a$, $b$, and $c$ correspond to the three photons-$a,b$, and $c$, respectively. We consider polarization-independent, temporal three-photon interference effects, assuming perfect spatial coherence. In Figs.~\ref{fig2}(b) and \ref{fig3}(b), $l$ represents the optical path length traveled by a photon, and $\phi$ represent phases other than the dynamical phase, including those acquired through reflections, geometric phases etc. Thus, $l_{a1}$ represents the optical path length traversed by photon-$a$ in alternative 1, etc. For each optical path length, the corresponding travel time is denoted by $\tau = l/c$. Therefore, $\tau_{a1}$ represents the time taken for photon-$a$ in reaching detector $D_a$ in alternative 1, etc. We note that although there are evidently sixteen alternative interference pathways in the setups of Figures~\ref{fig2}(a) and \ref{fig3}(a), the three-fold coincidence detection with a sufficiently narrow detection time window ensures that only two alternatives actually survive. We also emphasize that these prototypical setups are sufficiently generic that they can adequately model any specific experimental realization of three-photon interference because each of the eight length parameters is independently tunable in these setups. In other words, the corresponding three-photon path diagrams shown in Figs.~\ref{fig2}(b) and \ref{fig3}(b) are completely generic and can be used to model any specific experimental setup because any two-alternative three-photon interference scenario can at most involve eight different tunable length parameters. In what follows, we present our theoretical formalism for modeling three-photon interference scenarios. 

\subsection{Cascaded parametric down conversion (CPDC)}

Given the eight different length parameters involved in CPDC-based experiments (Figure \ref{fig2}), we define three length parameters and one phase parameter as follows:
\begin{align}
	\Delta L &= l_{1}-l_{2} \notag\\ &= \biggl(\frac{l_{a1}}{2}+ \frac{l_{b1}+l_{c1}}{4} + l_{p1} \biggr) - \biggl(\frac{l_{a2}}{2}+\frac{l_{b2}+l_{c2}}{4} + l_{p2} \biggr)\notag\\
	\Delta L' &= l_{1}' - l_{2}' = \biggl(\frac{l_{a1}}{2}-\frac{l_{b1}+l_{c1}}{4}\biggr) - \biggl(\frac{l_{a2}}{2}-\frac{l_{b2}+l_{c2}}{4}\biggr) \notag\\
	\Delta L'' &= l_{1}'' - l_{2}'' = \biggl(\frac{l_{b1}-l_{c1}}{2}\biggr) - \biggl(\frac{l_{b2}-l_{c2}}{2}\biggr)\notag\\
	\Delta \phi &= \phi_{1} - \phi_{2} = (\phi_{a1}+\phi_{b1}+\phi_{c1}+\phi_{p1}) \notag \\ & \qquad\qquad\qquad\qquad -(\phi_{a2}+\phi_{b2}+\phi_{c2}+\phi_{p2}) \label{def-CPDC}
\end{align}
Here, $l_{1(2)}$ denotes the three-photon path length, while $l_{1(2)}'$ and $l_{1(2)}''$ denote the three-photon path asymmetry lengths in alternatives 1(2). We refer to $\Delta L$ as the three-photon path-length difference while  $\Delta L'$ and $\Delta L''$ as the three-photon path-asymmetry-length difference parameters. The corresponding travel times are denoted as $\Delta \tau$, $\Delta \tau'$, and $\Delta \tau''$.

\subsection{Third order parametric down conversion (TOPDC)}

Given the eight different length parameters involved in TOPDC-based experiments (Figure \ref{fig3}), we define three length parameters and one phase parameter as follows:
\begin{align}
	\Delta L &= l_{1}-l_{2} = \biggl( \frac{l_{a1}+l_{b1}+l_{c1}}{3} + l_{p1} \biggr) \notag \\ &- \biggl( \frac{l_{a2}+l_{b2}+l_{c2}}{3} + l_{p2} \biggr)\notag \\
	\Delta L' &= l_{1}' - l_{2}' = (l_{a1}-l_{b1}) - (l_{a2}-l_{b2}) \notag \\
	\Delta L'' &= l_{1}'' - l_{2}'' = (l_{a1}-l_{c1}) - (l_{a2}-l_{c2})\notag\\
	\Delta \phi &= \phi_{1} - \phi_{2} = (\phi_{a1}+\phi_{b1}+\phi_{c1}+\phi_{p1}) \notag \\ & \qquad\qquad\qquad\qquad -(\phi_{a2}+\phi_{b2}+\phi_{c2}+\phi_{p2})  \label{def-TOPDC}
\end{align}
The above set of definitions is only one of three choices that works. The other two choices correspond to those obtained through a cyclic exchange of labels corresponding to the three photons. As shown in Appendix \ref{subsec:symmetry}, the equivalence of the three definitions manifests itself in having the same expressions for the coincidence count rate. Therefore, one can choose either of the three sets of definitions.

\section{Calculating the three-photon coincidence count rate} \label{sec:Theory}

In this section, we calculate the three-photon coincidence count rate corresponding to the three-photon path diagrams depicted in Figs.~\ref{fig2} and \ref{fig3}. We write the positive-frequency parts of the complex analytic field operators at detectors $D_{a}$, $D_{b}$, and  $D_{c}$ as
\begin{subequations}
	\begin{align}
		\hat{E_{a}}^{(+)}(t) &= k_{a1} \hat{E}_{a1}^{(+)}(t - \tau_{a1}) + k_{a2} \hat{E}_{a2}^{(+)}(t - \tau_{a2})\\
		\hat{E_{b}}^{(+)}(t) &= k_{b1} \hat{E}_{b1}^{(+)}(t - \tau_{b1}) + k_{b2} \hat{E}_{b2}^{(+)}(t - \tau_{b2})\\
		\hat{E_{c}}^{(+)}(t) &= k_{c1} \hat{E}_{c1}^{(+)}(t - \tau_{c1}) + k_{c2} \hat{E}_{c2}^{(+)}(t - \tau_{c2}),
	\end{align}\label{3fields}
\end{subequations}
where $\hat{E}_{a1}^{(+)}(t - \tau_{a1}) $ is the field of photon-$a$ at time $t - \tau_{a1}$ and $k_{a1}$ is the overall transmission coefficient of photon-$a$ to detector  $D_a$, etc. We note that the field $\hat{E_{a}}^{(+)}(t)$ at detector $D_a$ is the sum of the fields arriving at the detector through alternatives 1 and 2, etc.  The exact form of $	\hat{E_{a}}^{(+)}(t)$ can be written as: 
\begin{multline}
	\hat{E_{a}}^{(+)}(t) = k_{a1} e^{i\phi_{a1}} \int d\omega_{a} \hat{a}_{a1}(\omega_{a}) e^{-i\omega_{a} (t- \tau_{a1})}\\ \hspace{15mm}+ k_{a2} e^{i\phi_{a2}} \int d\omega_{a} \hat{a}_{a2}(\omega_{a}) e^{-i\omega_{a} (t- \tau_{a2})},\label{quantised}
\end{multline}
%
%
%
%
%
etc. In a three-photon interference experiment, the complete three-photon state $|\psi_{3p}\rangle$ is given by the coherent superposition of the three-photon states in alternatives 1 and 2, that is $|\psi_{3p}\rangle=|\psi_{3p1}\rangle +|\psi_{3p2}\rangle$. The three-photon coincidence count rate can be defined as the probability per (unit time)$^3$ that a photon is detected at time $t$, the second one at $t+t'$ and the third one at $t+t''$, and it can be written as $R_{abc}(t,t+t',t+t'')\equiv \alpha_{a}\alpha_{b}\alpha_{c}\langle\langle\psi_{3p}|\hat{E}_{a}^{(-)}(t) \hat{E}_{b}^{(-)}(t+t') \hat{E}_{c}^{(-)}(t+t'')\hat{E}_{c}^{(+)}(t+t'') \hat{E}_{b}^{(+)}(t+t')\hat{E}_{a}^{(+)}(t) |\psi_{3p}\rangle\rangle_e$ \cite{glauber1963pr, glauber1963pr2}, where $\alpha_a$, $\alpha_b$, and $\alpha_c$ are the detection efficiencies of the three detectors.
\begin{align}
	R_{abc}(& t,t+t', t+t'') =  \notag\\
	&R_{abc}^{11}(t,t+t',t+t'')+ R_{abc}^{22}(t,t+t',t+t'') \notag\\&+ R_{abc}^{12}(t,t+t',t+t'') + R_{abc}^{21}(t,t+t',t+t'')\label{rabc}
\end{align}
where
\begin{align}
	&R_{abc}^{12}(t,t+t',t+t'')=\alpha_{a}\alpha_{b}\alpha_{c} k_{a1}^{*}k_{b1}^{*}k_{c1}^{*}k_{a2}k_{b2}k_{c2} \notag \\ &\times \langle\langle\psi_{3p1}|\hat{E}_{a1}^{(-)}(t - \tau_{a1})  \hat{E}_{b1}^{(-)}(t+t' - \tau_{b1}) \notag\\ &\times \hat{E}_{c1}^{(-)}(t+t'' - \tau_{c1}) \hat{E}_{a2}^{(+)}(t - \tau_{a2}) \notag\\ &\times \hat{E}_{b2}^{(+)}(t+t' - \tau_{b2}) \hat{E}_{c2}^{(+)}(t+t'' - \tau_{c2})|\psi_{3p2}\rangle\rangle.\label{rabc-12}
\end{align}
Here, $\langle\cdots\rangle$ denotes the ensemble average. In most three-photon experiments, one does not measure the instantaneous coincidence rate $R_{abc}^{12}(t,t+t',t+t'')$. Instead, one measures the time-averaged coincidence count rate, averaged over the photon collection time $T_{\rm pc}$ and the two coincidence time-windows $T_{\rm ci1}$, and $T_{\rm ci2}$. Therefore, the time-averaged three-photon coincidence count rate  $R_{abc}$ can be obtained by integrating $R_{abc}^{12}(t,t+t',t+t'')$ with respect to $t$ over $T_{\rm pc}$, and
with respect to $t'$ and $t''$ over $T_{\rm ci1}$ and $T_{\rm ci2}$, respectively. In most experiments, the coincidence time-windows span a few nanoseconds, which is much longer than the inverse frequency bandwidth of down-converted photons, typically of the order of picoseconds. The photon collection time $T_{\rm pc}$ is usually a few seconds and is much longer than the inverse frequency bandwidth of the pump field, typically of the order of microseconds. Therefore, we perform the above time-averaging in the limit $T_{\rm pc}\to \infty$, $T_{\rm ci1}\to \infty$, and $T_{\rm ci2}\to \infty$  to get 
\begin{align}
	R_{abc}^{12}&=\langle R_{abc}^{12}(t,t+t', t+t'')\rangle_{t,t',t''} \notag \\
	&\propto \int_{-\infty}^{\infty}\int_{-\infty}^{\infty}\int_{-\infty}^{\infty} R_{abc}^{12}(t,t+t', t+t'') dtdt'dt''.\label{rabc-av}
\end{align}
and therefore, the time-averaged thee-photon coincidence count rate can be written as
\begin{align}
	R_{abc}=R_{abc}^{11}+ R_{abc}^{22}+ R_{abc}^{12}+R_{abc}^{22} \label{rabc-av-total}
\end{align}

\subsection{Cascaded parametric down conversion (CPDC)}\label{sec:cpdc}

The three-photon state $|\psi_{3p1}\rangle$ produced by CPDC in alternative 1 can be written as \cite{hong1985pra, franson1989prl, ou1989pra, grice1997pra,kim2000pra,jha2008pra,kulkarni2017josab, jha2010pra, borschevskaya2015lpl}:
\begin{multline}
	|\psi_{3p1}\rangle =  K_1\iiint d\omega_{a} d\omega_{b} d\omega_{c} V(\omega_{a}+\omega_{b}+\omega_{c}) \\ \times \Phi_{I}(\omega_{a}, \omega_{p}-\omega_{a}) \Phi_{II}(\omega_{b},\omega_{c}) \\ \times e^{-i(\omega_{p}\tau_{p1}+\phi_{p1})} |\omega_{a}\rangle_{a} |\omega_{b}\rangle_{b} |\omega_{c}\rangle_{c}.
	\label{psi3p}
\end{multline}
Here, $\omega_p$ is the frequency of the pump field, while $\omega_a$, $\omega_b$, and $\omega_c$ are the frequencies of the three photons, satisfying $\omega_p = \omega_a + \omega_b + \omega_c$. The term $V(\omega_a + \omega_b + \omega_c)$ denotes the pump spectral amplitude, $\Phi_{I}(\omega_a, \omega_p - \omega_a)$ and $\Phi_{II}(\omega_b, \omega_c)$ are the phase-matching functions of the first and second crystals, respectively. Furthermore, $\tau_{p1}$ is the travel time of the pump photon along the optical path $l_{p1}$, while $\phi_{p1}$ accounts for the accumulated phase, excluding the dynamical component. Let us first derive $\hat{E}_{a2}^{(+)}(t - \tau_{a2}) \hat{E}_{b2}^{(+)}(t+t' - \tau_{b2}) \hat{E}_{c2}^{(+)}(t+t'' - \tau_{c2})|\psi_{3p2}\rangle$. We substitute from Eqs.~(\ref{3fields}), (\ref{quantised}) into Eq.~(\ref{psi3p}) obtain
\begin{multline}
	\hat{E}_{a2}^{(+)}(t - \tau_{a2}) \hat{E}_{b2}^{(+)}(t+t' - \tau_{b2}) \hat{E}_{c2}^{(+)}(t+t'' - \tau_{c2})|\psi_{3p2}\rangle \\= K_2  e^{i(\phi_{a2} + \phi_{b2} + \phi_{c2} + \phi_{p2})} \iiint \iiint d\omega_{a}' d\omega_{b}' d\omega_{c}' \\ \times d\omega_{a} d\omega_{b} d\omega_{c} V(\omega_{p}) \Phi_{I}(\omega_{p},\omega_{p}-\omega_{a}) \Phi_{II}(\omega_{b},\omega_{c}) e^{i\omega_{p}\tau_{p2}} \\ \times e^{-i\omega_{a}'(t-\tau_{a2})} e^{-i\omega_{b}'(t+t'-\tau_{b2})} e^{-i\omega_{c}'(t+t''-\tau_{c2})} \\ \times \hat{a}_{a2}(\omega_{a}')  \hat{a}_{b2}(\omega_{b}') \hat{a}_{c2}(\omega_{c}') |\omega_{a}\rangle_{a2} |\omega_{b}\rangle_{b2} |\omega_{c}\rangle_{c2}.
	\label{E2}
\end{multline}
Now evaluating the annihilation operations and integrating over $\omega_{a}', \omega_{b}', \omega_{c}'$, we obtain
\begin{multline}
	\hat{E}_{a2}^{(+)}(t - \tau_{a2}) \hat{E}_{b2}^{(+)}(t+t' - \tau_{b2}) \hat{E}_{c2}^{(+)}(t+t'' - \tau_{c2})|\psi_{3p2}\rangle \\ = K_2 e^{i\phi_{2}} \iiint_{0}^{\infty} d\omega_{a} d\omega_{b} d\omega_{c}  V(\omega_{p}) \Phi_{I}(\omega_{a},\omega_{p}-\omega_{a}) \\ \times \Phi_{II}(\omega_{b},\omega_{c}) e^{i\omega_{p}\tau_{p2}} e^{-i\omega_{a}(t-\tau_{a2})} e^{-i\omega_{b}(t+t'-\tau_{b2})} \\ \times e^{-i\omega_{c}(t+t''-\tau_{c2})} |\mathrm{vac}\rangle_{a2} |\mathrm{vac}\rangle_{b2} |\mathrm{vac}\rangle_{c2},
	\label{E2next}
\end{multline}
where $\phi_{2}=\phi_{a2}+\phi_{b2}+\phi_{c2}+\phi_{p2}$. We now substitute
\begin{subequations}
	\begin{align}
		\omega_{p}&=\omega_{a}+\omega_{b}+\omega_{c}\\
		\omega'&= \omega_{a} - \omega_{b} - \omega_{c}\\
		\omega''&= \omega_{b} - \omega_{c}
	\end{align}
\end{subequations}
such that 
\begin{subequations}
	\begin{align}
		\omega_{a} &= \frac{\omega_{p}}{2}+\frac{\omega'}{2} \label{omega1} \\
		\omega_{b} &= \frac{\omega_{p}}{4} - \frac{\omega'}{4} + \frac{\omega''}{2} \label{omega2}\\
		\omega_{c} &= \frac{\omega_{p}}{4} - \frac{\omega'}{4} - \frac{\omega''}{2}
		\label{omega3}
	\end{align}
\end{subequations}
%
%
%
Here, $\omega_{p}, \omega_{a}, \omega_{b}$ and $\omega_{c}$ are centered around $\omega_{p0}, \omega_{a0}, \omega_{b0}$ and $\omega_{c0}$, with a frequency bandwidth of $\Delta\omega_{p}, \Delta\omega_{a}, \Delta\omega_{b}$ and $\Delta\omega_{c}$, respectively. \ Eq.~(\ref{E2next}) thus becomes
\begin{multline}
	\hat{E}_{a2}^{(+)}(t - \tau_{a2}) \hat{E}_{b2}^{(+)}(t+t' - \tau_{b2}) \hat{E}_{c2}^{(+)}(t+t'' - \tau_{c2})|\psi_{3p2}\rangle \\ = K_{2} e^{i\phi_{2}} \iiint_{-\infty}^{\infty} d\omega_{p} d\omega' d\omega'' V(\omega_{p}) \Phi_{I}(\omega_{p},\omega') \\ \times \Phi_{II}(\omega_{p},\omega',\omega'') e^{-i\omega' \bigl[ \bigl(\frac{t}{2}-\frac{t+t'+t+t''}{4}\bigr) - \bigl( \frac{\tau_{a2}}{2}-\frac{\tau_{b2} + \tau_{c2}}{4} \bigr) \bigr]} \\ \times e^{-i\omega_{p}\bigl[ \bigl(\frac{t}{2}+\frac{t+t'+t+t''}{4}\bigr) - \bigl( \frac{\tau_{a2}}{2}+\frac{\tau_{b2}+\tau_{c2}}{4} + \tau_{p2}\bigr)\bigr]} \\ \times e^{-i\omega'' \bigl[ \bigl( \frac{t+t'}{2} - \frac{t+t''}{2} \bigr) - \bigl( \frac{\tau_{b2}-\tau_{c2}}{2} \bigr) \bigr]} |\mathrm{vac}\rangle_{a2} |\mathrm{vac}\rangle_{b2} |\mathrm{vac}\rangle_{c2}.
\end{multline}
Replacing $ \bigl[ \bigl(t+\frac{t'+t''}{4}\bigr) - \bigl( \frac{\tau_{a2}}{2}+\frac{\tau_{b2}+\tau_{c2}}{4} +\tau_{p2} \bigr)\bigr]$,  $\bigl[ -\bigl(\frac{t'+t''}{4}\bigr) - \bigl( \frac{\tau_{a2}}{2}-\frac{\tau_{b2} + \tau_{c2}}{4} \bigr) \bigr]$, and $ \bigl[ \bigl( \frac{t'-t''}{2} \bigr) - \bigl( \frac{\tau_{b2}-\tau_{c2}}{2} \bigr) \bigr]$ with $(T - \tau_{2})$, $-(\frac{t'+t''}{4} + \tau_{2}')$, and $(\frac{t'-t''}{2} -\tau_{2}'')$, respectively, we get
\begin{multline}
	\hat{E}_{a2}^{(+)}(t - \tau_{a2}) \hat{E}_{b2}^{(+)}(t+t' - \tau_{b2}) \hat{E}_{c2}^{(+)}(t+t'' - \tau_{c2})|\psi_{3p2}\rangle \\ = K_{2} e^{i\phi_{2}} \iiint_{-\infty}^{\infty} d\omega_{p} d\omega' d\omega'' V(\omega_{p}) \Phi_{I}(\omega_{p},\omega') \\ \times \Phi_{II}(\omega_{p},\omega', \omega'') e^{-i\omega_{p}(T - \tau_{2})} e^{i\omega'(\frac{t'+t''}{4} + \tau_{2}')} \\ \times e^{-i\omega''  (\frac{t'-t''}{2} -\tau_{2}'')}  |\mathrm{vac}\rangle_{a2} |\mathrm{vac}\rangle_{b2} |\mathrm{vac}\rangle_{c2}.
	\label{3photon-prob-amp}
\end{multline}
%
%
We have $\omega_{p0}= (\omega_{a0}+\omega_{b0}+\omega_{c0})$. We substitute $ \omega_p =  \omega_{p0} +\bar{\omega}_{p}, \quad \omega' =  \omega_0' +\bar{\omega}', \quad \text{and} \quad \omega'' =\omega_0'' +\bar{\omega}''$ in Eq.~(\ref{3photon-prob-amp}), where $\omega'_{0}=(\omega_{a0}-\omega_{b0}-\omega_{c0}), \omega''_{0}=(\omega_{b0}-\omega_{c0})$. We assume that the frequency bandwidth of the pump field is narrow such that the phase-matching function remains nearly constant within $ (\omega_{p0} - \Delta\omega_p/2, \omega_{p0} + \Delta\omega_p/2)$. This way the above integral can be written as a product of two separate integrals, one containing only the pump field and the other one containing the phase-matching function. Finally, using Eqs.~(\ref{rabc-12}), (\ref{rabc-av}), and (\ref{3photon-prob-amp}), we obtain the following expression for $R_{abc}^{12}$, which is a product of two integrals separated in $\Delta\tau$ and ($\Delta\tau',\Delta\tau'$). 
\begin{multline}
	R_{abc}^{12}=\langle R_{abc}^{12}(t,t+t',t+t'') \rangle_{t,t',t''}\\ =  |c|^{2} K_{1}^{*} K_{2} e^{-i\Delta \phi} \int_{-\infty}^{\infty} d\bar{\omega}_{p} \langle |V(\omega_{p0}+\bar{\omega}_{p})|^{2} \rangle \\ \times e^{-i( \omega_{p0}+\bar{\omega}_{p})\Delta \tau} \iint_{-\infty}^{\infty} d\bar{\omega}' d\bar{\omega}''  \langle|\Phi_{I}(\omega'_{0}+\bar{\omega}')|^{2}\rangle \\ \times  \langle |\Phi_{II}(\omega'_{0}+\bar{\omega}',\omega''_{0}+\bar{\omega}'')|^{2} \rangle e^{-i(\omega_{0}'+\bar{\omega}')\Delta \tau'} \\ \times  e^{-i(\omega_{0}''+\bar{\omega}'')\Delta \tau''}
\end{multline}
where $|c|^2=\alpha_{a}\alpha_{b}\alpha_{c} k_{a1}^{*}k_{b1}^{*}k_{c1}^{*}k_{a2}k_{b2}k_{c2}$, $\tau_{1}-\tau_{2}=\Delta \tau$ ; $\tau_{1}'-\tau_{2}'=\Delta \tau'$ ; $\tau_{1}''-\tau_{2}''=\Delta \tau''$ and $\phi_{2}-\phi_{1}=\Delta\phi $. After rearranging the above equation we get,
\begin{multline}
	R_{abc}^{12}=\langle R_{abc}^{12}(t,t+t',t+t'') \rangle_{t,t',t''}\\ =  |c|^{2} K_{1}^{*} K_{2} e^{-i(\Delta \phi+\omega_{p0}\Delta\tau + \omega'_{0}\Delta\tau' + \omega''_{0}\Delta\tau'')} \\ \times \int_{-\infty}^{\infty} d\bar{\omega}_{p} \langle |V(\omega_{p0}+\bar{\omega}_{p})|^{2} \rangle e^{-i\bar{\omega}_{p}\Delta \tau} \iint_{-\infty}^{\infty} d\bar{\omega}' d\bar{\omega}'' \\ \times \langle|\Phi_{I}(\omega'_{0}+\bar{\omega}')|^{2}\rangle  \langle |\Phi_{II}(\omega'_{0}+\bar{\omega}',\omega''_{0}+\bar{\omega}'')|^{2} \rangle \\ \times e^{-i\bar{\omega}'\Delta \tau'}  e^{-i\bar{\omega}''\Delta \tau''}.\label{rabc12-final}
\end{multline}
Now, we substitute Eq.~(\ref{rabc12-final}) into Eq.~(\ref{rabc-av-total}) and obtain
\begin{multline}
	R_{abc} = |c|^{2} [|K_{1}|^{2} + |K_{2}|^{2} + |K_{1}| |K_{2}| \gamma(\Delta \tau) \\ \times \gamma'(\Delta \tau',\Delta \tau'') \cos(\omega_{p0}\Delta \tau + \omega_{0}'\Delta \tau' + \omega_{0}''\Delta \tau'' + \Delta \phi)].\label{rabc-final}
\end{multline}
where
\begin{subequations}
	\begin{multline}
		\gamma(\Delta \tau) = \int d\bar{\omega}_{p} \langle |V(\omega_{p0}+\bar{\omega}_{p})|^{2}\rangle e^{-i\bar{\omega}_{p}\Delta \tau}
	\end{multline}
	and
	\begin{multline}
		\gamma'(\Delta\tau',\Delta \tau'') = \iint d\bar{\omega}' d\bar{\omega}''  \langle|\Phi_{I}(\omega'_{0}+\bar{\omega}')|^{2}\rangle \\ \times  \langle |\Phi_{II}(\omega'_{0}+\bar{\omega}',\omega''_{0}+\bar{\omega}'')|^{2} \rangle e^{-i\bar{\omega}' \Delta \tau'} e^{-i\bar{\omega}'' \Delta \tau''} .
	\end{multline}
\end{subequations}
%
%
%
In situations in which the amplitudes of the two alternatives are equal, that is, $|cK_{1}|^{2}=|cK_{2}|^{2}=\frac{C}{2}$, we obtain
\begin{multline}
	R_{abc} = C [1 + \gamma(\Delta \tau) \gamma'(\Delta \tau',\Delta \tau'') \\ \times \cos(\omega_{p0}\Delta \tau + \omega_{0}'\Delta \tau' + \omega_{0}''\Delta \tau'' + \Delta \phi)],\label{CPDC-time}
\end{multline}
%
Finally, we write Eq.~(\ref{CPDC-time}) in terms of the length parameters defined in Eq.~(\ref{def-CPDC}) as $\Delta L = c\Delta \tau $ and so on. Thus the above equation can be written as
\begin{multline}
	R_{abc} = C [1 + \gamma(\Delta L)\gamma'(\Delta L',\Delta L'') \\ \times \cos(k_{p0}\Delta L + k_{0}'\Delta L' + k_{0}''\Delta L'' + \Delta \phi)]. \label{CPDC-length}
\end{multline}
Here $k_{p0}= \frac{\omega_{p0}}{c}=(k_{a0}+k_{b0}+k_{c0})$ is the mean vacuum wave-vector magnitude of the pump field, while $k_{0}'= \frac{\omega'_{0}}{c}=(k_{a0}-k_{b0}-k_{c0})$ and $k_{0}''= \frac{\omega''_{0}}{c}=(k_{b0}-k_{c0})$, where $k_{a0}$, $k_{b0}$ and $k_{c0}$ are the mean vacuum wave-vector magnitudes of photon-$a$, photon-$b$ and photon-$c$, respectively.

\subsection{Third-order parametric down conversion (TOPDC)}\label{sec:topdc}

The three-photon state $|\psi_{3p1}\rangle$ produced by $\chi^{(3)}$ TOPDC in alternative 1 can be written as
\begin{multline}
	|\psi_{3p1}\rangle = K_1 \iiint d\omega_{a} d\omega_{b} d\omega_{c} V(\omega_{a}+\omega_{b}+\omega_{c}) \\ \times\Phi(\omega_{a}, \omega_{b}, \omega_{c}) e^{-i(\omega_{p}\tau_{p1}+\phi_{p1})} |\omega_{a}\rangle_{a} |\omega_{b}\rangle_{b} |\omega_{c}\rangle_{c},
\end{multline}
where $\Phi(\omega_{a}, \omega_{b}, \omega_{c})$ is the phase-matching function for the $\chi^{(3)}$ crystal, and $V(\omega_{a}+\omega_{b}+\omega_{c})$ is the amplitude of the pump field where $\omega_{p}=\omega_{a}+\omega_{b}+\omega_{c}$. In this case, we have
\begin{multline}
	\hat{E}_{a2}^{(+)}(t - \tau_{a2}) \hat{E}_{b2}^{(+)}(t+t' - \tau_{b2}) \hat{E}_{c2}^{(+)}(t+t'' - \tau_{c2})|\psi_{3p2}\rangle \\ = K_2 e^{i(\phi_{a2} + \phi_{b2} + \phi_{c2} + \phi_{p2})} \iiint \iiint d\omega_{a}' \\ \ \ \times d\omega_{b}' d\omega_{c}' d\omega_{a} d\omega_{b} d\omega_{c} V(\omega_{p}) \Phi(\omega_{a},\omega_{b},\omega_{c}) e^{i\omega_{p}\tau_{p2}} \\  \ \times e^{-i\omega_{a}'(t-\tau_{a2})} e^{-i\omega_{b}'(t+t'-\tau_{b2})} e^{-i\omega_{c}'(t+t''-\tau_{c2})} \\  \ \ \times \hat{a}_{a2}(\omega_{a}') \hat{a}_{b2}(\omega_{b}') \hat{a}_{c2}(\omega_{c}') |\omega_{a}\rangle_{a2} |\omega_{b}\rangle_{b2} |\omega_{c}\rangle_{c2}.
\end{multline}
Now, evaluating the annihilation operations and integrating over $\omega_{a}', \omega_{b}', \omega_{c}'$, we obtain
\begin{multline}
	\hat{E}_{a2}^{(+)}(t- \tau_{a2}) \hat{E}_{b2}^{(+)}(t+t' - \tau_{b2}) \hat{E}_{c2}^{(+)}(t+t'' - \tau_{c2})|\psi_{3p2}\rangle \\ = K_{2} e^{i\phi_{2}} \iiint_{0}^{\infty} d\omega_{a} d\omega_{b} d\omega_{c}  V(\omega_{p}) \Phi(\omega_{a},\omega_{b},\omega_{c}) \\ \times e^{i\omega_{p}\tau_{p2}} e^{-i\omega_{a}(t-\tau_{a2})} e^{-i\omega_{b}(t+t'-\tau_{b2})} e^{-i\omega_{c}(t+t''-\tau_{c2})} \\ \times|\mathrm{vac}\rangle_{a2} |\mathrm{vac}\rangle_{b2} |\mathrm{vac}\rangle_{c2},
	\label{E2topdc}
\end{multline}
where $\phi_{2}=\phi_{a2}+\phi_{b2}+\phi_{c2}+\phi_{p2}$. We take,

\begin{subequations}
	\begin{align}
		\omega_{p}&=\omega_{a}+\omega_{b}+\omega_{c} \label{omgp}\\
		\omega'&=\frac{\omega_{a}+\omega_{c}}{2}-\omega_{b} \label{omgd1} \\
		\omega''&=\frac{\omega_{a}+\omega_{b}}{2}-\omega_{c} \label{omgd2}
	\end{align}
\end{subequations}
Thus,
\begin{subequations}
	\begin{align}
		\omega_{a}&=\frac{\omega_{p}}{3}+\frac{2\omega'}{3}+\frac{2\omega''}{3} \label{omg1}\\
		\omega_{b}&=\frac{\omega_{p}}{3}-\frac{2\omega'}{3} \label{omg2} \\
		\omega_{c}&=\frac{\omega_{p}}{3}-\frac{2\omega''}{3}. \label{omg3}
	\end{align}
\end{subequations}
Substituting the above equations into Eq.~(\ref{E2topdc}), we obtain
\begin{multline}
	\hat{E}_{a2}^{(+)}(t - \tau_{a2}) \hat{E}_{b2}^{(+)}(t+t' - \tau_{b2}) \hat{E}_{c2}^{(+)}(t+t'' - \tau_{c2})|\psi_{3p2}\rangle \\ = K_{2} e^{i\phi_{2}} \iiint_{0}^{\infty} d\omega_{p} d\omega' d\omega'' \Phi(\omega_{p},\omega',\omega'') V(\omega_{p}) \\ e^{-i\omega_{p} \bigl[ \bigl( \frac{t+t+t'+t+t''}{3} \bigr) - \bigl( \frac{\tau_{a2}+\tau_{b2}+\tau_{c2}}{3} + \tau_{p2} \bigr)\bigr]} \\ e^{-i\frac{2\omega'}{3}\bigl[(t-t-t') - (\tau_{a2} - \tau_{b2})  \bigr]} \\  e^{-i\frac{2\omega''}{3}\bigl[ (t-t-t'') - (\tau_{a2} - \tau_{c2}) \bigr]} |\mathrm{vac}\rangle_{a2} |\mathrm{vac}\rangle_{b2} |\mathrm{vac}\rangle_{c2}.
\end{multline}
Substituting $(\mathcal{T} - \tau_{2})$ for $ \bigl( t+\frac{t'+t''}{3} \bigr) - \bigl( \frac{\tau_{a2}+\tau_{b2}+\tau_{c2}}{3} + \tau_{p2} \bigr)$, 
$ -(t' + \tau_{2}')$ for $-t' - (\tau_{a2} - \tau_{b2})$, and  $ -(t'' + \tau_{2}'')$ for $-t'' - (\tau_{a2} - \tau_{c2}) $, we get
\begin{multline}
	\hat{E}_{a2}^{(+)}(t - \tau_{a2}) \hat{E}_{b2}^{(+)}(t+t' - \tau_{b2}) \hat{E}_{c2}^{(+)}(t+t'' - \tau_{c2})|\psi_{3p2}\rangle \\ = K_{2} e^{i\phi_{2}} \iiint_{0}^{\infty}  d\omega_{p} d\omega' d\omega'' V(\omega_{p}) e^{-i\omega_{p} (\mathcal{T} - \tau_{2})} \\ \times \Phi(\omega_{p},\omega',\omega'') e^{i\frac{2\omega'}{3}(t' + \tau_{2}')}  e^{i\frac{2\omega''}{3}(t'' + \tau_{2}'')]} \\ \times |\mathrm{vac}\rangle_{a2} |\mathrm{vac}\rangle_{b2} |\mathrm{vac}\rangle_{c2}.
	\label{appref}
\end{multline}
We assume that the pump field is centered at frequency \(\omega_{p0}=\omega_{a0}+\omega_{b0}+\omega_{c0}\) where $\omega_{a0}$, $\omega_{b0}$ and $\omega_{c0}$ are the central frequencies of  photon-$a$, photon-$b$ and photon-$c$ respectively. We consider $ \omega_p =  \omega_{p0} +\bar{\omega}_{p}, \quad 2\omega'/3 = (\omega_0' +\bar{\omega}'), \quad \text{and} \quad 2\omega''/3 = (\omega_0'' +\bar{\omega}'')$, where  $\omega_{0}'=\frac{\omega_{a0}+\omega_{c0}-\omega_{b0}}{3}$ and $\omega_{0}''=\frac{\omega_{a0}+\omega_{b0}-\omega_{c0}}{3}$. The frequency bandwidths of $2\omega'/3$ and $2\omega''/3$ are decided by $\Delta\omega_{a}$, $\Delta\omega_{b}$, and $\Delta\omega_{c}$. Now we substitute these in Eq.~(\ref{appref}) and extend the lower limits of integration to \(-\infty\). This is with the assumption that the frequency bandwidths of the pump and the down-converted photons are much smaller compared to their central frequencies.  Furthermore, we assume that the phase-matching function remains nearly constant over the spectral width of the pump field \((\omega_{p0} - \Delta\omega_p/2, \omega_{p0} + \Delta\omega_p/2)\). With these assumptions, we can write the above equation as a product of two separate integrals, that is,
\begin{multline}
	\hat{E}_{a2}^{(+)}(t - \tau_{a2}) \hat{E}_{b2}^{(+)}(t+t' - \tau_{b2}) \hat{E}_{c2}^{(+)}(t+t'' - \tau_{c2})|\psi_{3p2}\rangle \\ =  K_{2} e^{i\phi_{2}} \int_{-\infty}^{\infty} d\bar{\omega}_{p} V(\omega_{p0}+\bar{\omega}_{p}) e^{-i(\bar{\omega}_{p}+\omega_{p0})(\mathcal{T} -\tau_{2})} \\ \times \iint_{-\infty}^{\infty} d\bar{\omega}' d\bar{\omega}'' \Phi(\omega'_{0}+\bar{\omega}',\bar{\omega}''_{0}+\bar{\omega}'') e^{i(\omega_{0}' + \bar{\omega}')(t'+\tau_{2}')} \\ \times e^{i(\omega_{0}''+\bar{\omega}'')(t''+\tau_{2}'')}  |\mathrm{vac}\rangle_{a2} |\mathrm{vac}\rangle_{b2} |\mathrm{vac}\rangle_{c2}\label{topodc-amp}
\end{multline}
Using Eqs.~(\ref{rabc-12}), (\ref{rabc-av}), and (\ref{topodc-amp}), we obtain the following expression for $R_{abc}^{12}$:
\begin{multline}
	R_{abc}^{12} = \langle R_{abc}(t,t+t',t+t'')\rangle_{t_{\text{collect}},t_{\text{coin1}},t_{\text{coin2}}} \\= |c|^{2} K_{1}^{*} K_{2} e^{-i\Delta \phi} \int_{-\infty}^{\infty} d\bar{\omega}_{p} \langle |V(\omega_{p0}+\bar{\omega}_{p})|^{2} \rangle \\ \times e^{-i(\omega_{p0}+\bar{\omega}_{p})\Delta \tau} \iint_{-\infty}^{\infty} d\bar{\omega}' d\bar{\omega}'' e^{-i(\omega_{0}'+\bar{\omega}') \Delta \tau'} \\ \times  e^{-i(\omega_{0}''+\bar{\omega}'')\Delta \tau''} \langle |\Phi(\omega'_{0}+\bar{\omega}',\omega''_{0}+\bar{\omega}'')|^{2} \rangle 
\end{multline}
where $\tau_{1}-\tau_{2}=\Delta \tau$ ; $\tau_{1}'-\tau_{2}'=\Delta \tau'$ ; $\tau_{1}''-\tau_{2}''=\Delta \tau''$ and $\phi_{2}-\phi_{1}=\Delta\phi$. After rearranging, we get
\begin{multline}
	R_{abc}^{12} = \langle R_{abc}(t,t+t',t+t'')\rangle_{t_{\text{collect}},t_{\text{coin1}},t_{\text{coin2}}} \\= |c|^{2} K_{1}^{*} K_{2} e^{-i(\Delta \phi+\omega_{p0}\Delta\tau + \omega'_{0}\Delta\tau' + \omega''_{0}\Delta\tau'')} \\ \times \int_{-\infty}^{\infty} d\bar{\omega}_{p} \langle |V(\omega_{p0}+\bar{\omega}_{p})|^{2} \rangle e^{-i\bar{\omega}_{p}\Delta \tau} \\ \times \iint_{-\infty}^{\infty} d\bar{\omega}' d\bar{\omega}'' e^{-i\bar{\omega}' \Delta \tau'}  e^{-i\bar{\omega}''\Delta \tau''} \\ \times \langle |\Phi(\omega'_{0}+\bar{\omega}',\omega''_{0}+\bar{\omega}'')|^{2} \rangle 
\end{multline}
The time-averaged coincidence count rate, as defined in Eq.~(\ref{rabc-av-total}), now becomes
\begin{multline}
	R_{abc} = |c|^{2} [|K_{1}|^{2} + |K_{2}|^{2} + |K_{1}| |K_{2}| \gamma(\Delta \tau) \\ \times \gamma'(\Delta \tau',\Delta \tau'') \cos(\omega_{p0}\Delta \tau + \omega_{0}'\Delta \tau' + \omega_{0}''\Delta \tau'' + \Delta \phi)] \label{topdc-final}
\end{multline}
where
\begin{subequations}
	\begin{multline}
		\gamma(\Delta \tau)= \int \langle |V(\omega_{p0}+\bar{\omega}_{p})|^{2}\rangle e^{-i\bar{\omega}_{p}\Delta \tau} d\bar{\omega}_{p}
	\end{multline}
	and
	\begin{multline}
		\gamma'(\Delta\tau',\Delta \tau'') =\iint  \langle |\Phi(\omega'_{0}+\bar{\omega}',\omega''_{0}+\bar{\omega}'')|^{2} \rangle \\ \times e^{-i\bar{\omega}' \Delta \tau'} e^{-i\bar{\omega}'' \Delta \tau''} d\bar{\omega}' d\bar{\omega}''.
	\end{multline}
\end{subequations}
Here, $\gamma(\Delta \tau)\gamma'(\Delta \tau', \Delta \tau'')$ is the time-averaged degree of three-photon coherence. In situations in which the amplitudes of the two alternatives are equal, that is, $|cK_{1}|^{2}=|cK_{2}|^{2}=C/2$, Eq.~(\ref{topdc-final}) becomes
\begin{multline}
	R_{abc} = C [1 + \gamma(\Delta \tau) \gamma'(\Delta \tau',\Delta \tau'') \\ \cos(\omega_{p0}\Delta \tau + \omega_{0}'\Delta \tau' + \omega_{0}''\Delta \tau'' + \Delta \phi)].\label{TOPDC-time}
\end{multline}
Finally, in terms of the length parameters as defined in Eq.~(\ref{def-TOPDC}), the above equation can be written as
\begin{multline}
	R_{abc} = C [1 + \gamma(\Delta L)\gamma'(\Delta L',\Delta L'') \\ \cos(k_{p0}\Delta L + k_{0}'\Delta L' + k_{0}''\Delta L'' + \Delta \phi)].\label{TOPDC-length}
\end{multline}
Here, $k_{p0}$ is the mean vacuum wave vector magnitude of the pump wave, while $k_{0}' =\frac{\omega_{0}'}{c}= (\frac{k_{a0} + k_{c0}-k_{b0}}{3})$ and $k_{0}'' =\frac{\omega_{0}''}{c}= (\frac{k_{a0} + k_{b0}-k_{c0}}{3})$, where $k_{a0}$, $k_{b0}$, and $k_{c0}$ correspond to the mean vacuum wave vector magnitudes of photon-$a$, photon-$b$, and photon-$c$ fields, respectively. 
%
%

\section{Categorizing the three-photon interference effects}\label{sec:effects}

Equations (\ref{CPDC-time}) and (\ref{CPDC-length}) are the three-photon interference formulas for the CPDC process while Eqs.~(\ref{TOPDC-time}) and (\ref{TOPDC-length}) are the formulas for the TOPDC process. These formulas are in terms of only three length parameters. The degree of three-photon coherence $\gamma(\Delta L)\gamma'(\Delta L',\Delta L'')$ also factorizes in terms of these parameters. This essentially means that any three photon interference effect involving CPDC and TOPDC can be completely categorized in term of the variations of either the three-photon path-length difference ($\Delta L$) or the three-photon path-symmetry-length difference parameters ($\Delta L'$ and $\Delta L''$). Therefore, the three-photon interference effects can be divided into three distinct categories:

\subsection{Category I: Effects due to the variations of $\Delta \phi$}\label{subsec:case1}

In our analysis to see the three-photon effect due to the variations of $\Delta \phi$, we write Eqs.~(\ref{CPDC-length}) and (\ref{TOPDC-length}) in the limit $\Delta L=0, \Delta L'=0, \Delta L'' =0$ to obtain
\begin{align}
	R_{abc} = C [1 +  \cos( \Delta \phi)].
\end{align}
Thus, we obtain a cosine variation in the three-photon coincidence count rate with respect to the phase $\Delta \phi$, which is consistent with the observations reported in Refs. \cite{agne2017prl} and \cite{menssen2017prl}. For instance, Agne et al.\cite{agne2017prl} experimentally realized a generalized Franson setup involving a CPDC-based triplet source and recorded cosine interference fringes in the three-photon coincidence count rate with a visibility of ($92.7 \pm 4.6$)\%. 

\subsection{Category II: effects due to the variations of $\Delta L$}\label{subsec:case2}

In order to see the three-photon effects exclusively due to the variations of $\Delta L$, we write Eqs.~(\ref{CPDC-length}) and (\ref{TOPDC-length}) in the limit $\Delta L'=\Delta L''=0$. In this case $\gamma'(\Delta L', \Delta L'')=1$, and therefore, both equations ~(\ref{CPDC-length}) and (\ref{TOPDC-length}) take the following form:
\begin{align}
	R_{abc} = C [1 + \gamma(\Delta L) \cos(k_{p0}\Delta L + \Delta \phi)]
\end{align} 
We find that as a function of $\Delta L$, the three-photon coincidence count rate exhibits an  interference pattern with a fringe period equal to the central wavelength of the pump field $\lambda_{p0}=2\pi/k_{p0}$, and that the fringes get washed out once $\Delta L$ become much longer compared to width of $\gamma(\Delta L)$, that is, the coherence length of the pump field. Thus, we note that $\Delta L$ plays the same role in three-photon interference as does the optical path-length difference in one-photon interference and two-photon path-length difference in two-photon interference.   

\subsection{Category III: effects due to the variations of $\Delta L'$ and $\Delta L''$}\label{subsec:case3}

In order to highlight the effects due to the variations of three-photon path-asymmetry-length difference parameters ($\Delta L'$ and $\Delta L''$), we write Eqs.~(\ref{CPDC-length}) and (\ref{TOPDC-length}) in the limit $\Delta L=0$, $k_0'=0$, and $k_0''=0$. In this case $\gamma(\Delta L)=1$, and therefore, Eqs.~(\ref{CPDC-length}) and (\ref{TOPDC-length}) can be written as
\begin{align}
	R_{abc} = C [1 + \gamma'(\Delta L', \Delta L'') \cos(\Delta \phi)].
\end{align}
We find that as a function of $\Delta L'$ or $\Delta L''$, there are no interference fringes. Instead, what one observes is the profile of the degree of three-photon coherence function $\gamma'(\Delta L', \Delta L'')$. When $\Delta\phi=0$, the profile is observed in the form of a hump whereas when $\Delta\phi=\pi$, the profile appears in the form of a dip. Thus we see that the three-photon path-asymmetry-length difference parameter has no analogy with single-photon interference. However, it is analogous to how the two-photon path-asymmetry-length difference parameter works in two-photon interference. To this end, we note that in the context of a two-photon interference setup in which the signal and idler photons are mixed at a beam splitter, the degree of two-photon coherence as a function of the two-photon path-asymmetry-length difference appears as a dip, which is referred to as the HOM effect \cite{hong1987prl}. While this effect is often interpreted in terms of the bunching of two entangled photons at a beam splitter, such an interpretation comes into question in the famous postponed compensation experiment \cite{pittman1996prl}, wherein the two photons do not arrive at the beam-splitter at the same time.  Subsequently, other studies proposed that the HOM effect is more accurately explained as arising from the variation of the degree of two-photon coherence with respect to the path-asymmetry-length difference between the photons, which can occur even without photon mixing at a beam splitter \cite{jha2008pra, kulkarni2017josab}. We refer to the three-photon counterpart of this effect as the \enquote*{HOM-like effect}. Therefore, the observation of how the degree of three-photon coherence changes as a function of $\Delta L'$ and $\Delta L''$ at a fixed $\Delta L$ can be referred to as the three-photon HOM-like effect. This effect can appear in the form of a dip or a hump depending on the fixed value of $\Delta\phi$. We note that in the three-photon case, the HOM-like effect is dictated by two separate path-asymmetry-length difference parameters ($\Delta L'$ and $\Delta L''$), whereas in the two-photon case, it is dictated by only one such parameter. This implies that one can expect a wider variety of HOM-like effects in the three-photon case compared to the two-photon case.

\section{Summary and Outlook}\label{conclusion}

In summary, we present a formalism for describing temporal three-photon interference in terms of a generalized version of Dirac's dictum: \enquote{a three-photon interferes with itself}, which, as pointed out by Glauber \cite{glauber2006rmp}, can also be understood in terms of the interference of three-photon probability amplitudes.  We consider prototypical two-alternative three-photon interference setups involving CPDC- and TOPDC-based three-photon sources and show that in spite of the presence of eight tunable length parameters in the setups, there are only three independent length parameters that actually govern the interference. The first parameter is the three-photon path-length difference, which has a direct analog in the one-photon and two-photon interference. The other two length parameters quantify the path-length asymmetry. The variation of the three-photon coherence with respect to these two parameters results in a wider variety of inherently nonclassical HOM-like effects than those observed in two-photon interference, which involves only one path asymmetry length parameter. Our work can provide a theoretical basis for existing and future three-photon experiments exploring the rich and complex quantum correlations associated with three-particle entanglement. In addition, our study may also inspire the development of novel quantum protocols based on the unique HOM-like effects of three-photon interference that are absent in the two-photon counterpart.

\section*{Acknowledgements} \label{sec:acknowledgements}
We acknowledge financial support from Science and Engineering Research Board through grants (STR/2021/000035, VJR/2021/000012) $\&$ (CRG/2022/003070) and the Department of Science and Technology, Ministry of Science and Technology, India through the grant (DST/ICPS/QuST/Theme-1/2019). We also acknowledge financial support through the National Quantum Mission (NQM) of the Department of Science and Technology, Government of India.  N.S. thanks the Prime Minister’s Research Fellowship (PMRF), Ministry of Education, Government of India for financial support. G.K acknowledges the initiation research grant received from IIT Ropar.

\bibliographystyle{quantum}
\bibliography{three-photon}

\onecolumn
\appendix
\section{Three alternative ways of defining length parameters for TOPDC}

\label{subsec:symmetry}

In case of TOPDC, there are three possible sets of definitions of the length parameters, each giving the same expression for the three-photon coincidence count rate.

\

\noindent \textbf{Choice 1:}

As already given in Eqs.~\ref{omgp}, \ref{omgd1}, \ref{omgd2}, the first set of definitions involves choosing  $\omega'=\frac{\omega_{a}+\omega_{c}}{2}-\omega_{b} ; \omega''=\frac{\omega_{a}+\omega_{b}}{2}-\omega_{c}$. Correspondingly, the time parameters are defined as
\begin{align*}
	\tau_{2}'&=(\tau_{a2} - \tau_{b2}) ,  &\tau_{2}''=(\tau_{a2} - \tau_{c2});\\
	\tau_{1}'&=(\tau_{a1} - \tau_{b1}) ,  &\tau_{1}''=(\tau_{a1} - \tau_{c1});\\
	\text{and} \quad \Delta\tau' &= \tau_{1}'-\tau_{2}',  &\Delta\tau'' = \tau_{1}''-\tau_{2}''.
\end{align*}
And, the length parameters [as mentioned in Eq.~(\ref{def-TOPDC})] are, 
\begin{align}
	\Delta L &= l_{1}-l_{2} \notag \\ &= \biggl( \frac{l_{a1}+l_{b1}+l_{c1}}{3} + l_{p1} \biggr) - \biggl( \frac{l_{a2}+l_{b2}+l_{c2}}{3} + l_{p2} \biggr)\notag \\
	\Delta L' &= l_{1}' - l_{2}' = (l_{a1}-l_{b1}) - (l_{a2}-l_{b2}) \notag \\
	\Delta L'' &= l_{1}'' - l_{2}'' = (l_{a1}-l_{c1}) - (l_{a2}-l_{c2}).
\end{align}

\

\noindent \textbf{Choice 2:}

The second choice is: $\omega'=\frac{\omega_{b}+\omega_{c}}{2}-\omega_{a} ; \omega''=\frac{\omega_{b}+\omega_{a}}{2}-\omega_{c}$. The corresponding time parameters are given by:
\begin{align*}
	\tau_{2}'&=(\tau_{b2} - \tau_{a2});  &\tau_{2}''=(\tau_{b2} - \tau_{c2});\\
	\tau_{1}'&=(\tau_{b1} - \tau_{a1});  &\tau_{1}''=(\tau_{b1} - \tau_{c1});\\
	\text{and} \quad  \Delta\tau' &= \tau_{1}'-\tau_{2}',  &\Delta\tau'' = \tau_{1}''-\tau_{2}''.
\end{align*}
Subsequently, the length parameters take the form
\begin{align}
	\Delta L &= l_{1}-l_{2} \notag \\ &= \biggl( \frac{l_{a1}+l_{b1}+l_{c1}}{3} + l_{p1} \biggr) - \biggl( \frac{l_{a2}+l_{b2}+l_{c2}}{3} + l_{p2} \biggr)\notag \\
	\Delta L' &= l_{1}' - l_{2}' = (l_{b1}-l_{a1}) - (l_{b2}-l_{a2}) \notag \\
	\Delta L'' &= l_{1}'' - l_{2}'' = (l_{b1}-l_{c1}) - (l_{b2}-l_{c2}).
\end{align}

\noindent \textbf{Choice 3:}

The third choice is: $\omega'=\frac{\omega_{c}+\omega_{a}}{2}-\omega_{b} ; \omega''=\frac{\omega_{c}+\omega_{b}}{2}-\omega_{a}$. The corresponding time parameters are given by:
\begin{align*}
	\tau_{2}'&=(\tau_{c2} - \tau_{b2}) , &\tau_{2}''=(\tau_{c2} - \tau_{a2});\\
	\tau_{1}'&=(\tau_{c1} - \tau_{b1}) , &\tau_{1}''=(\tau_{c1} - \tau_{a1});\\
	\text{and} \quad  \Delta\tau' &= \tau_{1}'-\tau_{2}',  &\Delta\tau'' = \tau_{1}''-\tau_{2}''.
\end{align*}
Subsequently, the length parameters take the form
\begin{align}
	\Delta L &= l_{1}-l_{2} \notag \\ &= \biggl( \frac{l_{a1}+l_{b1}+l_{c1}}{3} + l_{p1} \biggr) - \biggl( \frac{l_{a2}+l_{b2}+l_{c2}}{3} + l_{p2} \biggr)\notag \\
	\Delta L' &= l_{1}' - l_{2}' = (l_{c1}-l_{b1}) - (l_{c2}-l_{b2}) \notag \\
	\Delta L'' &= l_{1}'' - l_{2}'' = (l_{c1}-l_{a1}) - (l_{c2}-l_{c2}).
\end{align}

\end{document}